# Research on Improved U-net Based Remote Sensing Image Segmentation Algorithm

Qiming Yang*

Engineering Research Center of Post Big Data Technology and Application of Jiangsu Province

Nanjing University of Posts and Telecommunications,

Jiangsu, China

Corresponding author:1716106991@qq.com

Zixin Wang

Digital Media

University of Washington

Seattle，The US

2722573328@qq.com

Shinan Liu

College of Engineering

Northeastern University

Boston, USA

Liu.shin@northeastern.edu

Zizheng Li

National Higher School of Advanced Techniques (ENSTA)

Paris, France

lizizheng2010@gmail.com

*Abstract*—In recent years, although U-Net network has made significant progress in the field of image segmentation, it still faces performance bottlenecks in remote sensing image segmentation. In this paper, we innovatively propose to introduce SimAM and CBAM attention mechanism in U-Net, and the experimental results show that after adding SimAM and CBAM modules alone, the model improves 17.41% and 12.23% in MIoU, and the Mpa and Accuracy are also significantly improved. And after fusing the two, the model performance jumps up to 19.11% in MIoU, and the Mpa and Accuracy are also improved by 16.38% and 14.8% respectively, showing excellent segmentation accuracy and visual effect with strong generalization ability and robustness. This study opens up a new path for remote sensing image segmentation technology and has important reference value for algorithm selection and improvement.

*Keywords- Remote sensing images；  U-Net network；semantic segmentation；  attention mechanism*

## I. INTRODUCTION

Remote sensing image segmentation technology plays an important role in urban planning and management, and requires efficient and accurate algorithms to adapt to the challenges of diverse urban architecture. Such algorithms cannot only provide accurate spatial information support to optimize urban layout and development trend prediction, but also play a key role in environmental monitoring and disaster assessment. Therefore, it is crucial to continuously promote the innovation and optimization of remote sensing image segmentation technology for the construction of smart cities and the sustainable development of society.

Nan Zhou [1] et al. proposed a multi-scale attention segmentation network (MsASNet), combined with channel attention mechanism, for automated remote sensing image recognition and achieved high accuracy. The model can effectively extract landslide information in real time, which is expected to be applied to the prevention and control of geological disasters. Jionghui Jiang [2] et al. proposed a deep learning semantic segmentation algorithm based on the dual-channel attention mechanism (DCAM), which utilizes the U-Net to combine RGB and near-infrared (NIR) images, and extracts features through the convolutional block attention module and the self-attention module, respectively, to achieve multi modal image feature complementation and association. Experiments show that the DCAM algorithm significantly improves segmentation accuracy, edge quality, and object integrity on the GID-15 dataset. Hua Zhang [3] et al. proposed an improved U-Net model based on migration learning for solving semantic segmentation problems in high-resolution remote sensing images. The model combines the symmetric encoder-decoder structure and multi-scale fusion technique to achieve good results on the ISPRS Vaihingen dataset, with significant improvement in MIoU and automobile class IoU, respectively, compared with the traditional U-Net model. Aiming at the noise and segmentation difficulties of multi-scale remote sensing images, Danying Liu [4] et al. proposed a U-net based segmentation algorithm through feature extraction, denoising and similarity function construction in order to enhance the edge features and optimize the segmentation details. Experiments show that the algorithm is able to accurately locate buildings and retain complete edges, which improves the segmentation effect. Jiaju Li[5] et al. designed a PSE-UNet model for semantic segmentation of hyperspectral remote sensing images by introducing a non-overlapping sliding window strategy combined with a judgment mechanism, and by combining the PCA, the attention mechanism, and the UNet, which outperforms other algorithms.

Liu Shiqi [6] in aircraft remote sensing image segmentation, the proposed U-Net model performs optimally, with mIOU up to $0.8432$ and Acc as high as $0.9971$, which effectively handles the remote sensing information and realizes accurate segmentation of aircraft, and is better than FCN and U-Net++, which is suitable for image segmentation needs in the fields of town construction, automatic driving, and so on. Liu Mingwei [7] and others proposed a deep learning-based semantic segmentation method for remote sensing images, which combines channel attention and multi-scale feature fusion to improve the classification accuracy and realize refined segmentation. Experiments show that the model reaches $90.88\%$ accuracy in surface coverage classification in southern China, and significantly outperforms U-Net for road and building segmentation, demonstrating high accuracy and good generalization ability. Xingjian Gu[8] et al. propose the AESwin-UNet model, which combines the advantages of CNN and Transformer, and captures local and global features through hybrid U-architecture and jump connections. capturing local and global features. AESwin-UNet performs well on WHDLD and LoveDA datasets, verifying its efficiency in semantic segmentation tasks of remote sensing images. Yang Yang [9] et al. proposed an attention-based multi-scale maximal pooling dense network (DMAU-Net) for feature classification of high-resolution remote sensing imagery. The DMAU-Net is used for feature classification in Vaihingen and Potsdam datasets with excellent performance, improved feature classification accuracy, clear feature boundaries and region integrity, integrated maximum pooling and efficient channel attention module design.

Based on the above in-depth analysis of the attention mechanism module, an enhanced U-Net remote sensing image segmentation algorithm is innovatively proposed in this paper. The algorithm skillfully integrates two efficient attention mechanisms, SimAM and CBAM, in the classical U-Net network architecture, aiming to significantly improve the segmentation performance of remote sensing images by enhancing the network's ability to capture and utilize key features in the images. By combining the adaptive weighting mechanism of SimAM with the dual spatial and channel attention mechanism of CBAM, this algorithm not only enhances the network's attention to the important features, but also further optimizes the quality of the feature representations, which comprehensively improves the accuracy and robustness of remote sensing image segmentation.

## II. METHODOLOGY

### A. U-Net Network Architecture

UNet is a deep learning model mainly used for image segmentation tasks. Its architecture is designed as an encoder-decoder model which handles pixel-level segmentation tasks through symmetric down-sampling and up-sampling. The UNet model is characterized by its U-shaped structure and the introduction of hopping connections between the encoder and the decoder, which helps to improve the accuracy and detail retention of the segmentation. A diagram of the UNet network architecture is as follows Fig. 1 shows.

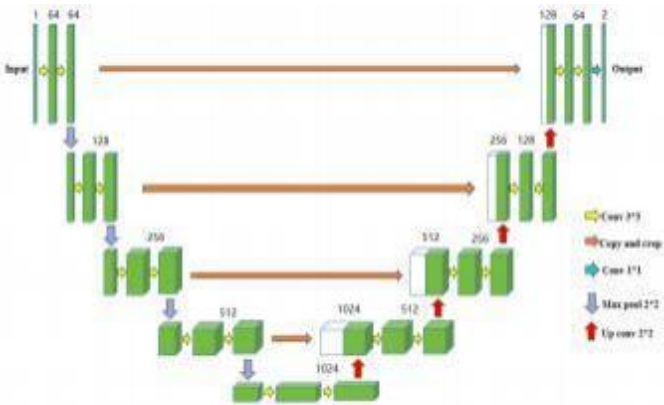

Fig. 1 U-net network architecture

The encoder consists of a convolutional layer and a pooling layer, the convolutional layer includes convolutional operations, batch normalization and activation functions for extracting local features and enhancing feature representation. The pooling layer, on the other hand, reduces the dimensionality of the feature map by means of maximum pooling, etc., which reduces the computation and parameters while retaining the important features and helps to extract the key features.

The jump connection layer is used in image segmentation tasks to convey different levels of feature information by adding jump connections between the decoder and encoder, which helps to retain more detailed information and improve accuracy and robustness. At the same time, this connection reduces data loss and plays an important role in model training.

The decoder part is responsible for mapping the feature maps extracted by the encoder back to the original image size and performing pixel-level classification. The components include an upsampling layer and a convolutional layer. The upsampling layer recovers the feature map size through an inverse convolution or interpolation operation, while the convolutional layer is used for feature fusion and refinement, which ultimately outputs pixel-level class predictions.

### B. Attention mechanism

1. CBAM module

CBAM (Convolutional Block Attention Module) is an attention mechanism module for enhancing the performance of Convolutional Neural Networks (CNN). Proposed by Sanghyun Woo and his team in 2018. CBAM introduces channel attention and spatial attention to enhance the model perceptual ability while improving the performance without increasing the complexity of the network. CBAM consists of two key parts, the channel attention module (CAM) and spatial attention module (SAM). Its specific working principle is shown in Fig. 2 below.

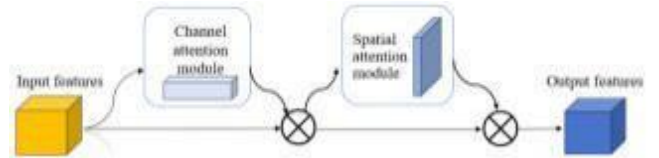

Fig. 2 CBAM module

The Channel Attention Module (CAM) is designed to enhance the feature expression of each channel. First, the maximum and average feature values on each channel are computed by global maximum pooling and global average pooling operations. Second, the pooled feature vectors are fed into a shared fully connected layer for learning the attention weights for each channel. Then, a Sigmoid activation function is applied to ensure that the attention weights are between 0 and 1. Finally, the obtained attention weights are multiplied with each channel of the original feature map to obtain an attention-weighted channel feature map.

Spatial Attention Module (SAM) is used to emphasize the importance of different locations in the image. The operation steps areas follows: firstly, global maximum pooling and global average pooling are performed on the channel feature maps to obtain two $H \times W \times 1$ feature maps; then these two feature maps are spliced in the channel dimension, and after convolution operation to reduce the dimension to 1 to obtain the spatial attention feature maps of $H \times W \times 1$; then the Sigmoid activation function is applied to generate the spatial attention weights; finally, the obtained attention weights are applied to the original feature map, weighting the features at each spatial location.

2. SimAM module

SimAM (Simple Attention Module) is a lightweight parameter-free attention mechanism designed to enhance the feature representation capability of convolutional neural networks. This mechanism enhances the important information in the feature map through an adaptive weighting mechanism without additional parameters. SimAM has low computational overhead along with performance enhancement.

The working principle of SimAM is mainly based on the following steps:

(1) Input feature map: a feature map with multiple channels, each containing a two-dimensional spatial feature (height and width), is received as input.

(2) Calculate channel mean and variance: for each channel, calculate the mean of all its pixels to reflect the overall intensity level of that channel. The variance of each channel is calculated to reflect the degree of variation of pixel values within that channel. The greater the variance, the greater the variation in pixel values within that channel.

(3) Calculate Attention Weights: the calculated mean and variance are used to generate attention weights. This weight indicates the importance of each pixel location in the feature map. The calculation of the attention weight is based on an optimized energy function that aims to find the importance of each neuron.

(4) Applying Attention Weights: the computed attention weights are applied to each pixel location in the original feature map to enhance the important information in the feature map through weighting operations.

III. EXPERIMENTAL RESULTS

A. Experimental design

Based on the dataset from the Headquarters of Geodesy and Cartography in Poland, we implemented a fine-grained data enhancement strategy by cutting the original images into subimages of 512x512 pixels, which successfully expanded the number of 40 images to 10,370 and significantly enriched the training samples. Subsequently, we scientifically divide the dataset into training and testing sets at a ratio of 9:1 to ensure adequate training and accurate testing. Further, we keep one original image as an independent validation set to enhance the model generalization capability. This well-planned data preprocessing and segmentation process not only optimizes the training efficiency, but also significantly improves the robustness of the model and its practical application.

Assessing the accuracy of remote sensing image segmentation algorithms requires careful selection of evaluation metrics. In this study, three widely recognized and representative metrics are used: accuracy (A), mean pixel accuracy (MPA) and mean intersection and merger ratio (MIoU). These metrics are able to quantify the degree of match between segmentation results and real labels, providing a basis for comprehensively evaluating the performance of the algorithm. Their specific calculation formulas are as follows.

$$A = \frac{TP + TN}{TP + TN + FP + FN} \quad (1)$$

$$MPA = \frac{1}{n+1} \sum_{i=0}^{n} \frac{TP}{FN} \quad (2)$$

$$MIoU = \frac{1}{n+1} \sum_{i=0}^{n} \frac{TP}{FN + FP + TP} \quad (3)$$

In these formulas, TP represents the number of samples whose true value is a positive category and is correctly predicted as a positive category (true example); FN represents the number of samples whose true value is a positive category but is incorrectly predicted as a negative category (false negative example); FP represents the number of samples whose true value is a negative category but is incorrectly predicted as a positive category (false positive example); and TN represents the number of samples whose true value is a negative category and is correctly predicted as a negative category (true negative example); Here $k + 1$ denotes the number of categories.

B. Analysis of results

In this paper, we verify the practical improvement effect of the attention mechanism module on the performance of the U-Net model through ablation experiments. The experiment gradually integrates two advanced attention mechanism modules, SimAM and CBAM, into the U-Net model, compares the performance under different configurations, and analyzes the contribution of each module to the model performance. The experimental results show that the SimAM module and CBAM module have obvious effects in improving the model segmentation accuracy and enhancing the feature extraction ability, which verifies the effectiveness of the improvement strategy.

Table 1 Comparison of experimental results

| Method | U-net | SimAM | CBAM | MIoU | Mpa | Accuracy |
|---|---|---|---|---|---|---|
| a | ✓ | x | x | 65.45 | 70.32 | 80.43 |
| b | ✓ | ✓ | x | 82.86 | 84.2 | 93.38 |
| c | ✓ | x | ✓ | 77.68 | 81.35 | 90.07 |
| d | ✓ | ✓ | ✓ | **84.56** | **86.7** | **95.23** |

From the data analysis in Table 1, we can clearly see the significant enhancement of the performance of the U-Net model in remote sensing image segmentation tasks by the attention mechanism. After the introduction of the attention mechanism SimAM module, compared with the baseline model, the model's MIoU is improved by 17.41 percentage points, Mpa by 13.88 percentage points, and Accuracy by 9.95 percentage points. While adding the attention mechanism CBAM module, the MIoU is only improved by 12.23 percentage points, while the model's Mpa and Accuracy are improved by 11.03 and 9.64 percentage points, respectively. In order to further improve the model's performance, this paper combines the strategies of the SimAM and CBAM modules to realize a leap in the comprehensive performance of the model, with the MIoU, Mpa and Accuracy are improved by 19.11, 16.38 and 14.8 percentage points, respectively. This innovation not only verifies the complementary nature of the two attention mechanisms, but also highlights the great potential and practical application value of this paper's method in improving model accuracy and robustness.

In order to comprehensively evaluate the performance of U-Net in the remote sensing image segmentation task, we selected remote sensing images in several different scenes and accurately segmented these images. Subsequently, we selected three representative images from these scenes and analyzed them by comparing the segmentation results of different networks on this image. Where (A) denotes the original image; (B) denotes the U-net; (C) denotes the U-net adding the CBAM module; (D) denotes the U-net adding the SimAM module; and (E) denotes the U-net adding the SimAM module and the CBAM module; in the comparison image, red represents the correctly recognized building image element, blue represents the correctly recognized water image element, yellow represents the correctly recognized woodland image element, and green represents the correctly recognized road image element.

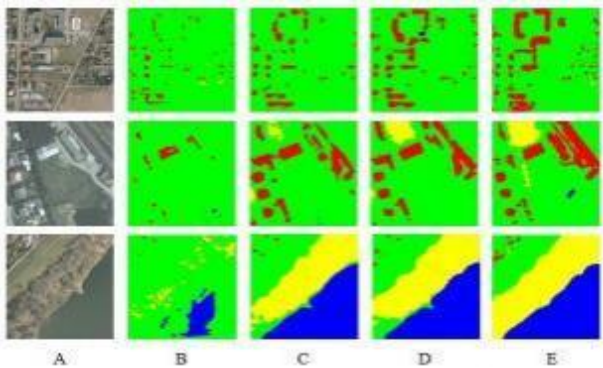

Fig. 3 Comparison of segmentation results

A closer look at the segmentation effect graph comparison in Fig. 3 shows that the U-Net network is significantly under-represented in recognizing the red (buildings) and blue (water) pixels, highlighting its limitation inaccurately segmenting these key features, and signaling the room for improving the overall accuracy. However, by introducing the SimAM and CBAM attention mechanisms into the U-Net architecture, the number of red, blue, and yellow (woodland) pixels are significantly improved, demonstrating the effectiveness of the attention mechanism in enhancing the feature extraction and classification capabilities of the model and optimizing the segmentation accuracy. It is especially remarkable that the parallel integration of SimAM and CBAM modules in the U-Net network exceeds the enhancement of a single attentional mechanism, and not only dramatically improves the recognition accuracy of key features, such as buildings, water and woodland, but also significantly increases the number of correctly recognized pixels, realizing the refinement of segmentation effects and a leap in accuracy, thus demonstrating the most excellent performance and results in the task of remote sensing image segmentation. This has resulted in the most excellent performance and results in remote sensing image segmentation tasks.

## IV. CONCLUSIONS

After in-depth analysis and experimental validation, it is clear that although U-Net has basic capabilities in remote sensing image segmentation, it is insufficient to capture the details of complex scenes, especially in the segmentation of buildings and water areas, which is a bottleneck. In order to breakthrough this limitation, we innovatively integrate SimAM and CBAM attention mechanisms into U-Net, and significantly improve the segmentation performance by dynamically adjusting the feature weights and multi-dimensional attention optimization. Experimental results show that our method significantly outperforms the original U-Net in key metrics such as MIoU, Mpa, and Accuracy, especially MIoU is improved by nearly 20%, and it shows good generalization and robustness. In the future, we will deepen the research on attention mechanism and network architecture design, and explore more efficient and intelligent remote sensing image segmentation models by combining advanced techniques such as self-attention and graph neural network, and emphasize on model interpretability and lightweight design to achieve a perfect balance between performance and efficiency.